# A Multimodal Hyperspectral and Laser Speckle Imaging Platform for Simultaneous Monitoring of Spinal Cord Oxygenation, Perfusion and Metabolism

Junda Wang[†a], Luca Giannoni[†*a], Ayse Gertrude Yenicelik[b], Eleni Giama[b], Frédéric Lange[a] and Ilias Tachtsidis[a]

[a] Department of Medical Physics and Biomedical Engineering, University College London, London, UK

[b] Department of Neuroinflammation, UCL Queen Square Institute of Neurology, London, UK

[†]These authors contributed equally.

*Corresponding author: Luca Giannoni; l.giannoni@ucl.ac.uk.

**Abstract**

Hyperspectral imaging (HSI) enables spatial mapping of haemoglobin oxygenation and cytochrome-c-oxidase (CCO) redox changes in exposed neural tissue. In this study, we present a technical upgrade of an existing HSI platform by the integration of laser speckle contrast imaging (LSCI), enabling contemporaneous assessment of oxygenation, oxidative metabolism, relative perfusion and a derived relative oxygen metabolic index in exposed spinal cord in a rat preclinical model. The upgraded system (hNIR+) uses a shared optical path and a single camera to acquire HSI and LSCI images from the same field of view using a sequential acquisition workflow. Hyperspectral images are acquired at 11 discrete wavelengths (600, 630, 665, 784, 800, 818, 835, 851, 868, 881, and 894 nm) to estimate changes in oxygenated haemoglobin ($HbO_2$), deoxygenated haemoglobin (HHb) and oxidized CCO (oxCCO). LSCI is implemented using 632.8-nm coherent illumination to generate relative blood-flow index (rBFI) maps. The LSCI subsystem was validated using a controlled blood-flow phantom, where the LSCI-derived inverse speckle contrast index showed a strong positive linear relationship with pump-controlled flow. The upgraded platform was then evaluated in proof-of-concept observations of the rat spinal cord under normoxia, hypoxia and recovery conditions. Representative multimodal maps and ROI-based time-course analysis showed clear dynamic responses in $HbO_2$, HHb, oxCCO and rBFI during oxygen challenges that were consistent with the expected physiological responses of the spinal cord during hypoxia, hyperoxia and subsequent recoveries. Overall, these results demonstrate that integrating LSCI into HSI extends the platform to multimodal and simultaneous monitoring of haemodynamic–metabolic coupling.

## 1. Introduction

Optical imaging is particularly attractive for preclinical and translational tissue monitoring because it can provide non-contact, label-free, and spatially resolved measurements over exposed biological surfaces [1,2]. Spectral imaging has been increasingly explored in biomedical and surgical applications because wavelength-dependent tissue reflectance contains information about tissue morphology, haemodynamics, and biochemical composition [3-6]. Hyperspectral imaging (HSI) extends conventional optical imaging by acquiring images across multiple spectral bands, allowing pixel-wise spectral analysis of endogenous chromophores. In the visible and near-infrared (NIR) range, the dominant physiological absorbers include oxygenated haemoglobin ($HbO_2$), deoxygenated haemoglobin (HHb), and cytochrome-c-oxidase (CCO). This makes HSI well suited for monitoring haemodynamic and metabolic changes in neural tissue [4, 5].

The hNIR system, developed by Giannoni *et al*. [7], was previously validated as a hyperspectral imaging platform for mapping haemoglobin and CCO concentration changes in exposed cerebral tissue, the first of its kind. In that framework, wavelength-dependent changes in optical attenuation were used to estimate relative changes in $HbO_2$, HHb, and oxidised CCO (oxCCO). The quantification of haemoglobin and oxCCO using near-infrared HSI has also been investigated through simulation studies, supporting the importance of wavelength selection and spectral fitting accuracy for resolving overlapping chromophore spectra [8]. The inclusion of oxCCO is particularly important because CCO is the terminal enzyme of the mitochondrial electron transport chain, and its oxidation state has been widely discussed as an optical indicator related to mitochondrial oxidative metabolism [9-11].

Despite these advantages, hNIR alone does not directly measure blood flow. This represents a significant limitation for physiological interpretation. For example, a decrease in $HbO_2$ may be caused by reduced perfusion, increased oxygen extraction, systemic desaturation, or a combination of these factors [11]. Similarly, a decrease in oxCCO may indicate impaired mitochondrial oxygen availability or altered metabolic demand, but without a perfusion-sensitive measurement it is difficult to determine whether the metabolic response is primarily driven by reduced vascular supply or cellular oxygen utilization [10].

Laser speckle contrast imaging (LSCI) provides a complementary approach for wide-field mapping of relative blood flow. When coherent laser light illuminates tissue, the scattered light forms a speckle pattern at the camera sensor. Motion of scattering particles, mainly red blood cells, causes temporal fluctuations and blurring of this speckle pattern during the camera exposure. The resulting reduction in speckle contrast can be used to estimate a relative blood-flow index [12-14]. Recent dynamic light scattering and LSCI studies have further clarified the relationship between speckle dynamics, model assumptions, exposure conditions, and flow estimation [15, 16]. These studies also highlight an important practical point: unless carefully calibrated, speckle-based measurements should generally be interpreted as relative perfusion or relative flow indices rather than absolute blood-flow velocity.

The combination of spectral imaging and LSCI has therefore become an attractive strategy for multimodal tissue assessment. HSI or multispectral imaging can provide oxygenation and chromophore information, while LSCI provides information on perfusion with high-temporal-resolution. Previous studies have demonstrated the feasibility of combining LSCI with snapshot HSI for real-time visualization of haemodynamics and tissue oxygenation [17, 18]. More recently, integrated LSCI and multispectral imaging has also been used for multimodal assessment of tissue viability, including blood-flow index, oxygen saturation, and tissue parameters related to metabolic rate [19-22]. These studies support the general concept that combining perfusion-sensitive and oxygenation-sensitive measurements can provide more physiologically informative data than either modality alone.

In the present study, we describe a technical upgrade of the existing hNIR imaging platform through the integration of an LSCI module, called hNIR+. The aim of the paper is to describe, characterise, validate, and demonstrate the added value of this upgrade. First, we describe the optical and technical add-ons required to incorporate LSCI into the existing hNIR system. Second, we introduce the acquisition and processing workflow used to generate maps of $HbO_2$, HHb, oxCCO and relative blood-flow index (rBFI). Third, we validate the LSCI component using a controlled blood-flow phantom experiment.

Finally, we demonstrate the feasibility of the upgraded multimodal platform in proof-of-concept rat spinal cord experiments under normoxia, hypoxia, hyperoxia and recovery conditions. The spinal cord is highly vulnerable to disturbances in blood supply, oxygen delivery, and mitochondrial metabolism. Spinal cord hypoperfusion or ischaemia may occur during traumatic injury, vascular compromise, surgical intervention, aortic procedures, systemic hypoxia, or reperfusion events, and can lead to irreversible neurological damage if not detected and managed promptly [23, 24]. Direct monitoring of spinal cord blood flow, oxygenation and metabolism can therefore be investigated as a means of detecting spinal cord ischaemia earlier than achieved by indirect electrophysiological warning signs [25]. Furthermore, the study of hypoxic-related disorders using rat spinal cord models, such as active and passive experimental autoimmune encephalomyelitis (EAE) models for multiple sclerosis (MS) [26, 27], could greatly benefit from such an instrument providing a comprehensive overview on tissue physiology in quasi-real time.

To our knowledge, this is the first study where HSI and LSCI are used simultaneously, in a single platform, to measure multiple physiological parameters including, $HbO_2$, HHb, relative blood flow, as well as oxCCO.

## 2. Materials and Methods

### *2.1 System configuration*

The multimodal imaging platform was developed as a technical upgrade of the previously reported hyperspectral near-infrared imaging system, hNIR, which was designed to map haemoglobin oxygenation and CCO redox changes in exposed neural tissue. The original hNIR subsystem has been described in detail elsewhere by Giannoni *et al*. [7]; therefore, the present section focuses primarily on the newly integrated LSCI module and the resulting multimodal configuration. This upgrade follows the broader concept of combining spectral imaging with flow-sensitive optical measurements to improve the physiological interpretation of tissue oxygenation, metabolism, and perfusion [20-22].

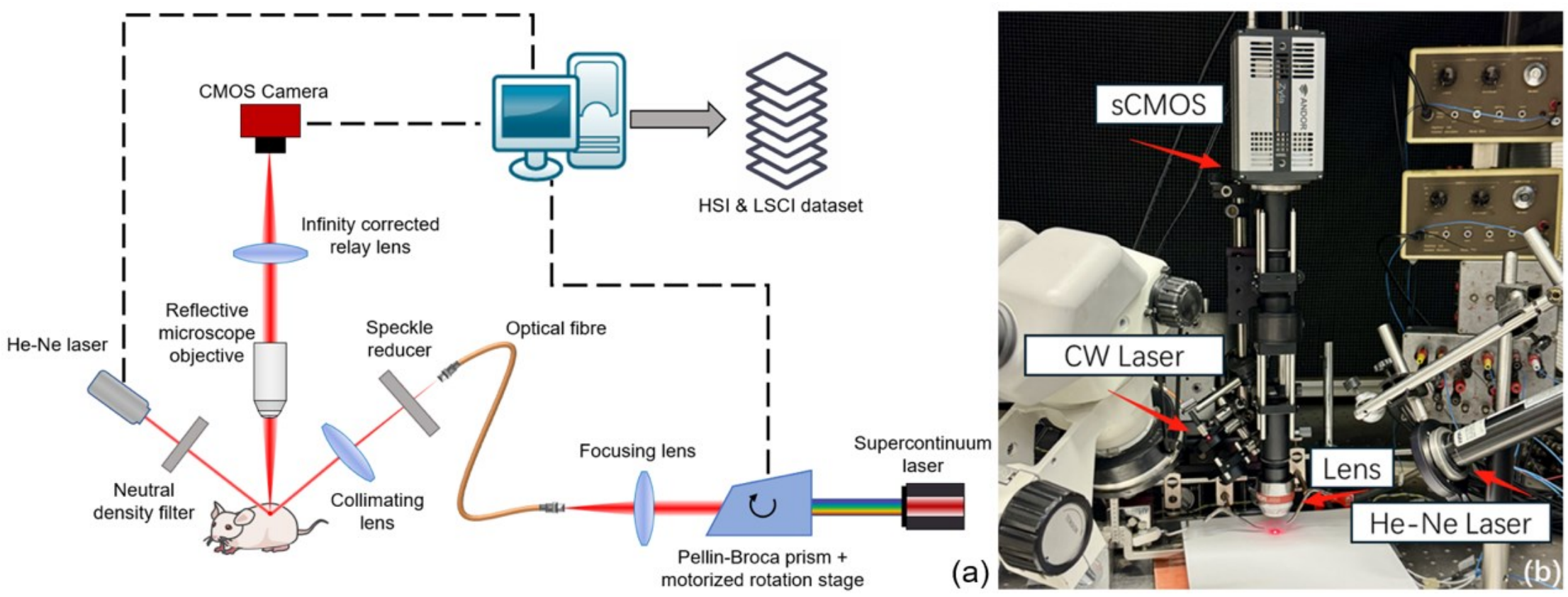

*Figure 1. Upgraded hNIR+ imaging platform. (a) Schematic diagram of the multimodal optical configuration. Hyperspectral NIR illumination was provided by a supercontinuum laser and motorized Pellin–Broca wavelength-selection unit, while laser speckle contrast imaging was performed using 632.8-nm He-Ne illumination. Both modalities shared the same collection path through a reflective objective, tube lens and sCMOS camera. (b) Photograph of the experimental setup showing the sCMOS camera, objective, illumination optics and He-Ne laser path used for LSCI acquisition.*

The upgraded system, named hNIR+ (Fig. 1), uses a shared imaging path and a single scientific CMOS camera to acquire both hyperspectral and laser speckle images from the same field of view. The detector is an Andor Zyla 5.5 sCMOS camera (Oxford

Instruments, UK), with a 2560 × 2160-pixel monochrome sensor and 16-bit dynamic range. Using the same detector for both modalities enables pixel-level comparison between HSI-derived chromophore maps and LSCI-derived blood-flow maps, while avoiding inter-camera parallax and reducing the need for post-acquisition spatial registration.

For hyperspectral acquisition, illumination is provided by a supercontinuum broadband laser source (Fianium WhiteLase SC440-40) coupled to a wavelength-selection unit based on a motorised Pellin–Broca prism (Thorlabs ADB-10). Narrowband illumination is delivered to the sample through optical fibres and illumination optics. Hyperspectral images are acquired at 11 discrete wavelengths: 600, 630, 665, 784, 800, 818, 835, 851, 868, 881, and 894 nm. These wavelengths were selected to cover the absorption features of $HbO_2$, HHb and oxCCO, following previous hNIR validation and simulation work on haemoglobin and oxCCO quantification [8,9]. The spectral bandwidth of each illumination band is approximately 8–11 nm full width at half maximum (FWHM) [7].

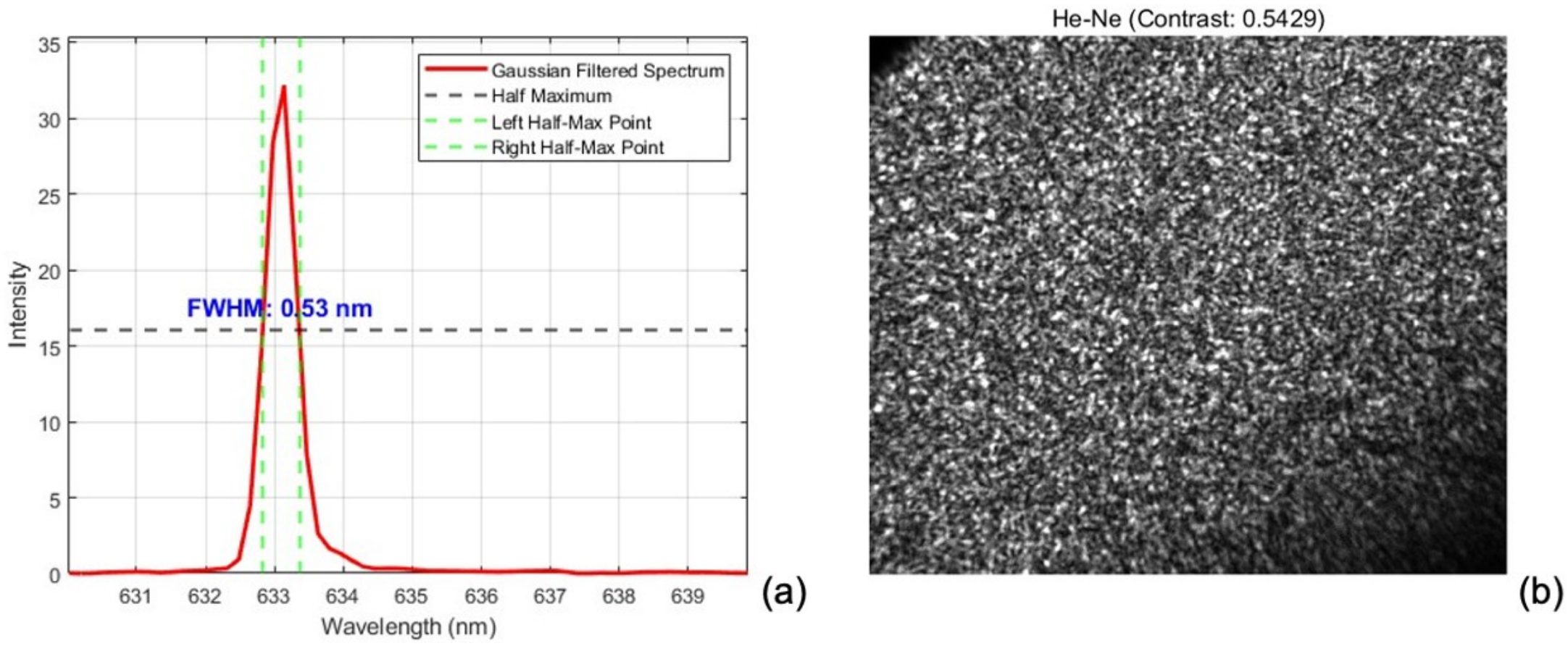

*Figure 2. Characterisation of the He–Ne laser used for LSCI acquisition. (a) Measured emission spectrum of the He–Ne laser, showing a narrow peak centred near 633 nm with a measured full width at half maximum (FWHM) of approximately 0.53 nm. (b) Representative static speckle image acquired under stationary conditions, showing clear speckle formation with a measured spatial speckle contrast of 0.5429.*

The LSCI module consists of a continuous-wave helium-neon (He–Ne) laser operating at 632.8 nm with a nominal output power of 4 mW (Model 1103-1669, Laser Drive Inc., USA). Before integration into the multimodal imaging platform, the laser source was characterised to verify its suitability for speckle contrast imaging (Fig. 2). Spectral characterisation showed a narrow emission peak centred close to 633 nm, with a measured FWHM of approximately 0.53 nm. This narrow spectral profile was considered suitable for producing coherent speckle illumination, as excessive laser spectral width has been shown to reduce speckle contrast and degrade LSCI performance [15, 28]. Static speckle measurements were also acquired under stationary conditions, producing a clearly resolved speckle pattern with a mean spatial speckle contrast of 0.5429. These measurements confirmed that the He-Ne source provided sufficient coherence and speckle visibility for subsequent LSCI acquisition.

The laser beam was expanded to provide approximately uniform illumination over the same tissue region imaged by the hNIR subsystem. Speckle images were acquired using millisecond-scale exposure times to capture flow-dependent speckle blurring while avoiding sensor saturation. During HSI acquisition, the LSCI laser was switched off; during LSCI acquisition, the hyperspectral illumination was switched off. This illumination-switching strategy prevented optical crosstalk between the two modalities, which is particularly important when combining coherent speckle illumination with spectral reflectance imaging.

Both modalities shared the same optical collection path. Light from the sample was collected using a 15× reflective objective (Thorlabs LMM-15X-P01) and relayed to the sCMOS camera through an infinity-corrected tube lens (Thorlabs TTL200-B). The reflective objective was selected to minimize chromatic aberration across the visible and near-infrared range and to maintain a consistent focal plane between hyperspectral and speckle imaging. The system provided a lateral spatial resolution of $<5$ μm, with a working distance of 23.8 mm and an effective field of view of approximately 1.1 mm × 0.9 mm [7]. Camera triggering, wavelength tuning, and illumination switching were controlled using a PC-based acquisition program.

*2.2 Multimodal acquisition workflow*

Multimodal imaging was performed using a sequential acquisition workflow. At each imaging time point, the system first acquired a full hyperspectral image stack at the predefined wavelengths, followed by laser speckle acquisition from the same field of view. This sequential workflow was selected to prevent optical interference between the hyperspectral and speckle illumination sources while maintaining spatial correspondence between oxygenation, metabolism, and perfusion measurements. Similar multimodal systems combining LSCI with hyperspectral or multispectral imaging have demonstrated the value of acquiring flow and oxygenation information from corresponding tissue regions [24,25].

Before each imaging session, dark and white reference images were collected for hyperspectral reflectance correction. The dark image was acquired with all illumination sources blocked, while the white reference image was acquired from a calibrated reflectance standard under the same hyperspectral illumination conditions. For LSCI, the laser illumination and exposure time were adjusted before acquisition to avoid camera saturation and maintain sufficient speckle contrast. These calibration steps were used to ensure consistency across repeated imaging sessions and to support comparison between baseline, hypoxia, hyperoxia, and recovery conditions.

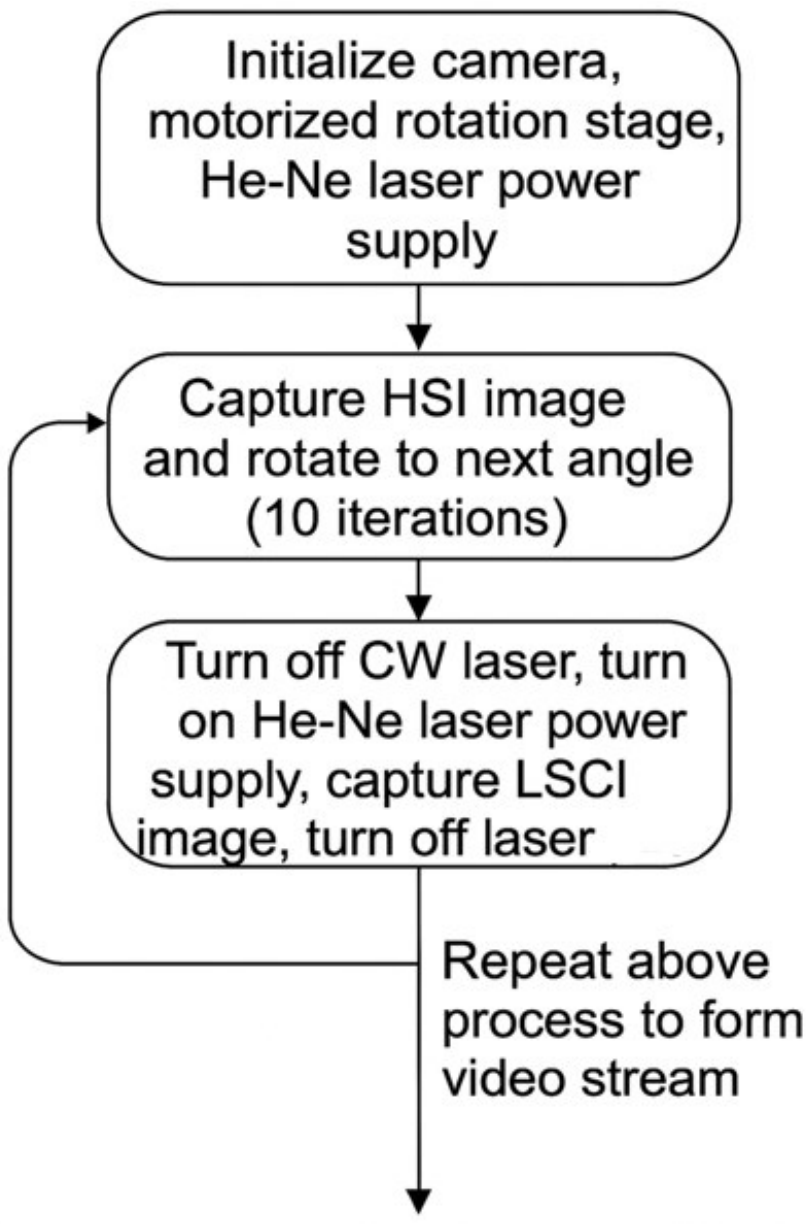

*Figure 3. Sequential multimodal acquisition workflow.*

During a typical acquisition cycle (Fig. 3), the Pellin–Broca prism was rotated sequentially to each calibrated spectral position, and one image was acquired at each wavelength. After completion of the 11-wavelength hyperspectral stack, the hyperspectral illumination was disabled, and the 632.8-nm coherent laser was activated for LSCI acquisition. This workflow enabled repeated multimodal measurements during both phantom and *in vivo* experiments while preserving the full dynamic range of the sCMOS detector. The duration of the complete acquisition of a single hyperspectral dataset, as well as the corresponding LSCI dataset (12 frames in total), depends on the exposure times for each modality (50 to 150 ms for HSI, 2 to 10 ms for LCSI), but typically ranged between 6 and 10 s (0.1-0.16 Hz).

*2.3 Blood-flow phantom experiment for LSCI validation*

A blood-flow phantom experiment was performed to characterize the LSCI subsystem and validate its sensitivity to controlled flow changes before *in vivo* imaging. The validation was based on the principle that motion of scattering particles causes temporal fluctuations and spatial blurring of the speckle pattern, leading to changes in speckle contrast that can be related to flow [12]. Controlled-flow phantom experiments are commonly used to evaluate whether a speckle-based system responds consistently to known changes in flow before being applied to biological tissue. Recent

LSCI phantom studies have also used embedded tubes, syringe-pump-driven flow, and controlled optical scattering conditions to assess the performance of speckle-based flow measurements [29].

The phantom consisted of a liquid blood-based optical mixture prepared using whole human blood. This mixture provided fully realistic optical scattering and absorption properties suitable for testing speckle formation and flow sensitivity. The phantom fluid was loaded into a syringe connected to transparent tubing of 600-μm diameter positioned within the imaging field of view. A programmable syringe pump was used to drive the fluid through the tubing at controlled flow settings. A close-up view of the tubing region positioned under the objective during LSCI acquisition is shown in Fig. 4.

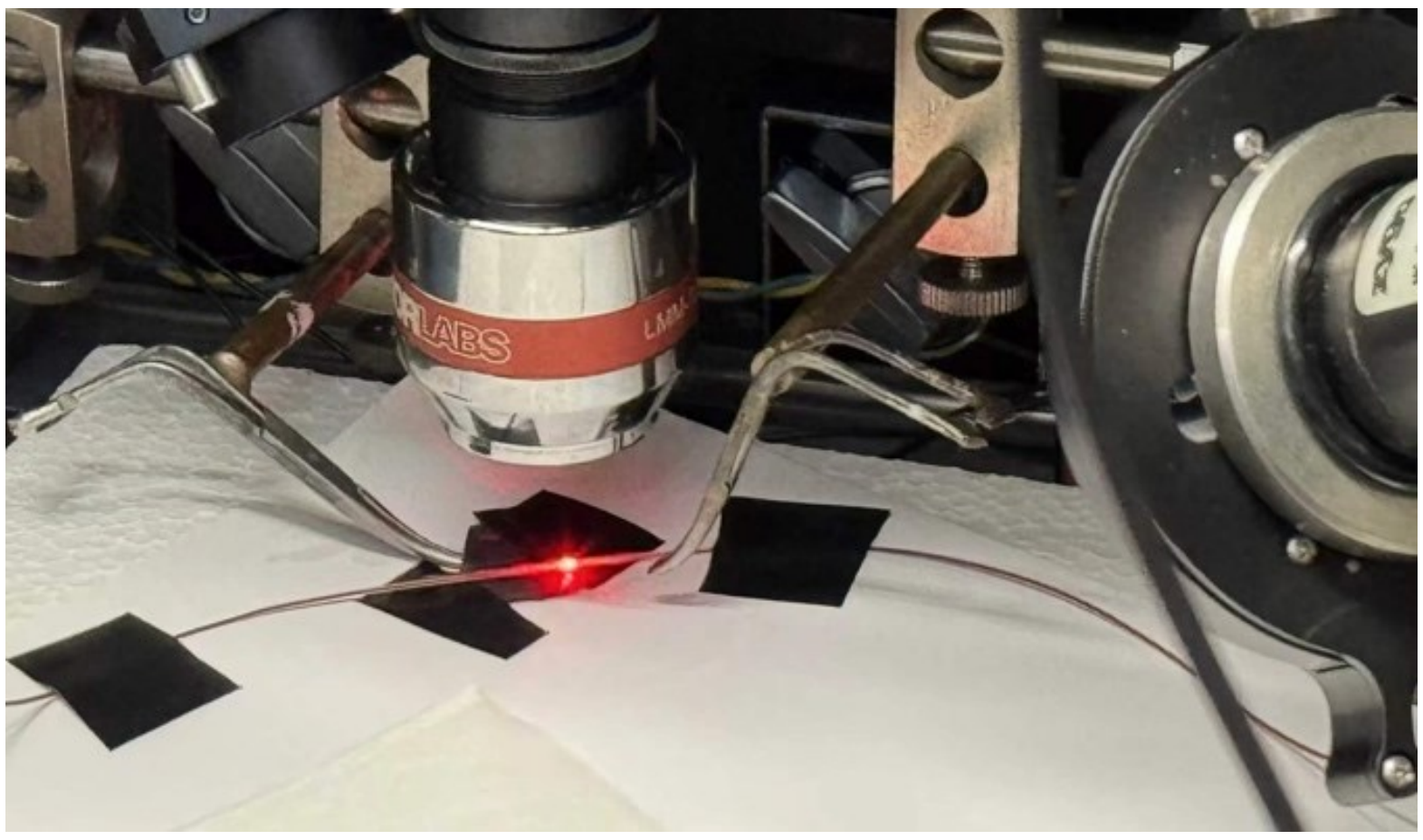

*Figure 4. Close-up photograph of the blood-flow phantom during LSCI validation*

The pump was programmed to modulate the flow in seven stages: 5, 15, 25, 35, 95, 25, and 5 arbitrary units, corresponding to increasing and decreasing relative flow conditions. At each flow stage, LSCI images were acquired using the same optical configuration and camera settings. A fixed region of interest over the tubing was selected for analysis, and speckle contrast was calculated from this region. The purpose of this experiment was to confirm that the LSCI-derived flow index changed consistently with the pump-controlled flow setting, supporting its use as a relative perfusion measurement in subsequent rat spinal cord experiments.

*2.4 Rat spinal cord proof-of-concept experiment*

The upgraded hNIR+ platform was evaluated in proof-of-concept rat spinal cord experiments under controlled regulation of the respiratory oxygen concentration. All animal experiments were performed in accordance with the UK Animals (Scientific Procedures) Act 1986. Three independent *in vivo* rat spinal cord imaging experiments were conducted in total (on three different animals). The experimental design aimed to test whether the upgraded system could capture co-registered changes in haemoglobin oxygenation, oxCCO, and rBFI during controlled respiratory oxygen challenges.

Anaesthesia was induced and maintained by intraperitoneal administration of a ketamine–xylazine mixture. The anaesthetic solution consisted of 65% Ketavet® containing ketamine at 100 mg/mL and 35% Sedaxylan® containing xylazine at 20 mg/mL. Animals received an induction dose of 1.5 μL/g body weight by intraperitoneal injection. During the experimental procedure, anaesthesia was maintained by continuous intraperitoneal infusion of the same ketamine–xylazine solution through an intraperitoneal catheter using a syringe pump at a rate of 1 μL/g body weight/h.

Animals were placed on a homoeothermic heating mat connected to a rectal temperature probe, and rectal temperature was continuously monitored and maintained at 36.5–37.0 °C throughout the experiment. A subcutaneous catheter was placed for continuous saline administration during the imaging procedure. Adequate depth of anaesthesia was assessed throughout the procedure by monitoring corneal reflexes and withdrawal reflexes to paw pinch. Respiratory rate was continuously monitored throughout the experiment.

Adult rats were positioned in a prone posture for imaging. An incision and dorsal laminectomy at the T12–L2 level were carried out to expose the thoracolumbar spinal cord for optical imaging. The exposed spinal cord surface was kept moist with sterile saline throughout the experiment to preserve tissue hydration and optical clarity. The imaging system was positioned above the exposed spinal cord, and the region of interest was centred within the camera field of view. The animal was secured to minimise motion, and a shielded, opaque enclosure was used to reduce external light contamination and electrical background noise during HSI acquisition and to improve speckle contrast during LSCI acquisition.

The respiratory challenge protocol was designed to induce controlled changes in oxygen delivery and perfusion so that LSCI-derived blood-flow information could be assessed alongside HSI-derived oxygenation and metabolic maps. The core protocol consisted of a baseline normoxic or room-air period, hyperoxia or hypoxic challenges, and recovery under room air at 21% $O_2$. Depending on the experimental run, hypoxic challenges were performed using inspired oxygen concentrations of 10%, or 7.5% $O_2$ balanced with nitrogen. Selected experiments included repeated hypoxia–recovery cycles or hyperoxia phases (~100% $O_2$)with varying duration depending on the subjects, and on experimental circumstances. The exact oxygen concentration and timing for the representative datasets are indicated in the results in Sec. 3.2 and 3.3.

Repeated multimodal acquisition cycles were collected throughout each experiment without changing the imaging geometry. The same exposed spinal cord region was monitored over time to evaluate whether changes in $HbO_2$, HHb, oxCCO, and rBFI followed physiologically consistent trends during oxygen challenges and recovery, as well as whether hNIR+ was able to correctly and performatively detect such changes.

*2.5 Hyperspectral image processing and chromophore mapping*

Hyperspectral image processing followed the previously reported hNIR analysis framework [7,8], with minor adaptations for the present multimodal dataset. Briefly, raw images acquired at the selected wavelengths were organized into hyperspectral image stacks (hypercubes), with two spatial dimensions and one spectral dimension. Where repeated acquisitions were available, images acquired under the same experimental condition were averaged over short time-windows to improve signal stability and reduce random sensor noise, consistent with the processing strategy used in the original system workflow. The full pipeline also included hypercube generation, reflectance extraction, spectral unmixing and optional post-processing [7, 8], but these steps are summarized here because the present paper focuses on the LSCI upgrade and multimodal validation rather than redeveloping the hNIR algorithm.

For each wavelength, raw pixel intensities were converted into calibrated reflectance values using dark and white reference images. The dark image was acquired with illumination blocked, while the white reference image was acquired from a calibrated reflectance standard under identical illumination conditions. The corrected reflectance was calculated as:

$$R(x,y,\lambda_i) = \frac{I_{\text{raw}}(x,y,\lambda_i) - I_{\text{dark}}(x,y,\lambda_i)}{I_{\text{white}}(x,y,\lambda_i) - I_{\text{dark}}(x,y,\lambda_i)} \tag{1}$$

where $I_{raw}(x,y,\lambda_i)$, $I_{dark}(x,y,\lambda_i)$, and $I_{white}(x,y,\lambda_i)$ represent the raw tissue image, dark image, and white reference image at wavelength $\lambda_i$, respectively. This correction reduced wavelength-dependent illumination variation and detector non-uniformity.

The corrected reflectance images were converted into optical density changes relative to the initial normoxic baseline period:

$$\Delta\, OD(x,y,\lambda_i,t) = -\log_{10}\left[\frac{R(x,y,\lambda_i,t)}{R(x,y,\lambda_i,\text{baseline})}\right] \tag{2}$$

Changes in $HbO_2$, HHb, and oxCCO were estimated using a modified Beer–Lambert law approach [7-9]. The measured optical density changes across the selected wavelengths were fitted as a linear combination of the extinction spectra of the target chromophores using least-squares fitting, after estimation of the optical pathlength via a Monte Carlo simulation platform developed in Giannoni *et al.* [8]. The resulting outputs were pixel-wise maps of $\Delta HbO_2$, $\Delta HHb$, and $\Delta oxCCO$, expressed as relative concentration changes from baseline rather than absolute concentrations. This HSI-derived oxygenation information was used for spatial visualization and ROI-based time-course analysis.

*2.6 LSCI data processing and relative blood-flow mapping*

Raw laser speckle images were processed to generate relative blood-flow maps. Speckle contrast was calculated using a spatial contrast method, consistent with recent dynamic light scattering and LSCI modelling studies [29, 30]. For each pixel, the local speckle contrast *K* was calculated within a sliding window as:

$$K(x, y, t) = \frac{\sigma_I(x, y, t)}{\langle I(x, y, t) \rangle} \tag{3}$$

where $\sigma_I(x, y, t)$ is the standard deviation of pixel intensity within the local window, and $\langle I(x, y, t) \rangle$ is the corresponding mean intensity. The window size was selected as a compromise between spatial resolution and noise reduction. A larger window improves contrast stability but reduces spatial detail, whereas a smaller window preserves spatial resolution but increases noise.

The speckle contrast maps were converted into a relative blood-flow index (rBFI) using the inverse contrast relationship:

$$rBFI(x, y, t) = \frac{1}{K(x, y, t)} \tag{4}$$

where $rBFI(x, y, t)$ represents the LSCI-derived relative blood-flow index. This inverse relationship was used to provide a relative measure of flow-dependent speckle blurring. For *in vivo* analysis, rBFI maps were normalised to the initial normoxic baseline period:

$$rBFI_{\mathrm{norm}}(x, y, t) = \frac{rBFI(x, y, t)}{rBFI_{\mathrm{baseline}}(x, y)} \tag{5}$$

Where $rBFI_{\mathrm{baseline}}(x, y)$was calculated as the mean rBFI value during the baseline period. The resulting normalised rBFI maps were used to visualise relative perfusion changes during normoxia, hypoxia, and recovery.

Because simplified speckle contrast analysis is affected by exposure time, static scattering, multiple scattering, motion artefacts, and calibration conditions, the resulting values were interpreted as relative perfusion changes rather than absolute blood-flow velocity [31]. For quantitative analysis, regions of interest were selected from the LSCI-derived rBFI maps, and ROI-averaged rBFI time courses were extracted.

*2.7 ROI and statistical analysis*

Regions of interest (ROI) were manually selected from anatomically and physiologically relevant areas of the rat spinal cord, particularly highlighting both intra- and extravascular compartments. Intravascular, vessel-dominated ROIs were chosen from regions showing clear vascular structures and strong flow contrast, while extravascular ROIs were selected from adjacent tissue regions with minimal large-vessel contribution. In particular, we identified an area in the grayscale images from which we selected the ROI where we were confident that image focus was stable and curvature effects from the tissue were not predominant interfering factors in the quantification (An example of this has been highlighted with a dotted lined in Fig. 6). The areas of confidence primarily excluded the edges of the images and areas downslope of the major vasculature, thus accounting for approximately 75% of the field of view in all the animal subjects.

For each ROI, mean values of the changes in concentration of $HbO_2$, HHb, oxCCO, as well as rBFI were extracted at each imaging time point. For the blood-flow phantom validation experiment, the relationship between the imposed pump flow setting and the LSCI-derived flow index was assessed using linear regression. For *in vivo* experiments, ROI-averaged time courses were compared across normoxia, hypoxia, hyperoxia, recovery, and severe hypoxia phases.

For blood-flow plots, vascular and extravascular ROIs were displayed using separate y-axes or separate panels where necessary, because the vascular blood-flow scale was larger and could obscure smaller extravascular changes. This allowed both vascular and extravascular blood-flow dynamics to be visualised more clearly. The same plotting strategy was also used when comparing oxygenation, metabolism, and perfusion trends across different experimental datasets.

## 3. Results

*3.1 Blood-flow phantom validation of the LSCI module*

The LSCI subsystem was first evaluated using a controlled blood-flow phantom before *in vivo* rat spinal cord imaging. This experiment was used to determine whether the newly integrated LSCI module could detect controlled changes in flow under the same optical configuration used for subsequent biological measurements.

Representative rBFI mapping of the syringe-pump phantom showed that the flow-sensitive signal was localized within the tubing region, yet inhomogeneities in the flux distribution are visibly present. These may be attributed to uneven distribution of the illumination (particularly between left and right side) due to the sharp, cylindrical curvature of the tube (much larger than the typical curvature of *in vivo* vasculature).

A fixed 100 × 100-pixel ROI was selected over the tubing for quantitative analysis, as shown in Fig. 5a. The rBFI map confirmed that the selected ROI captured the region in which pump-driven flow changes were expected to occur.

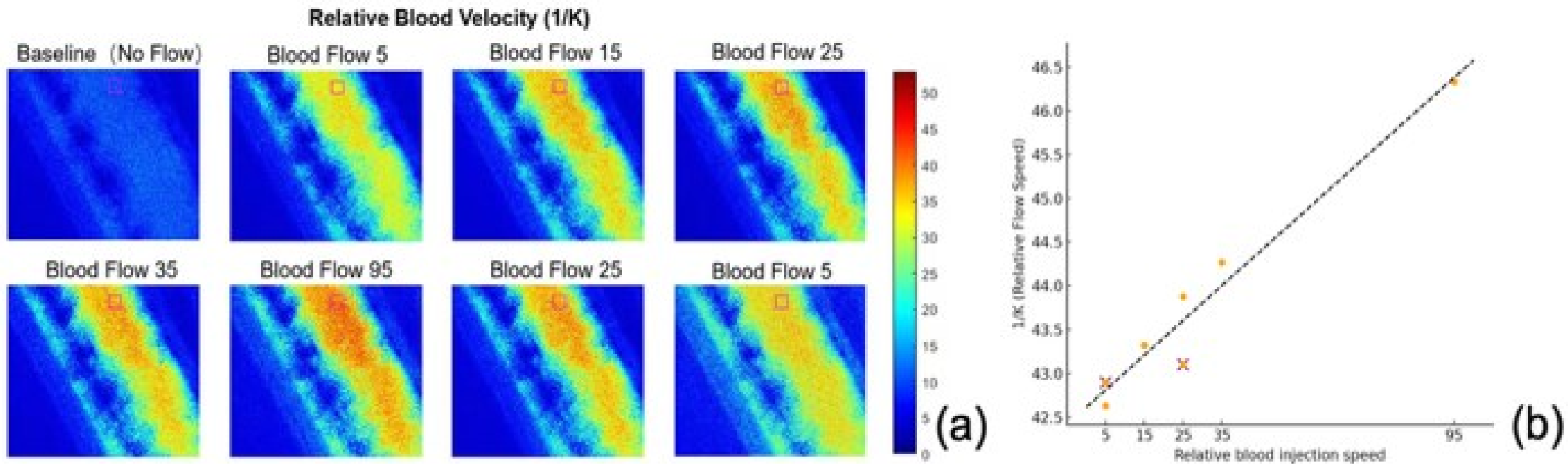

*Figure 5. Blood-flow phantom validation of the LSCI module. (a) Relative flow maps under no-flow and seven pump-flow stages. (b) Linear regression between pump setting and LSCI-derived 1/K.*

The pump speed was modulated through seven stages: 5, 15, 25, 35, 95, 25, and 5 arbitrary units. The LSCI-derived relative flow index was calculated as the inverse speckle contrast, 1/K. As the pump setting increased from 5 to 95 a.u., the measured 1/K value increased from 42.63 to 46.33. When the pump setting was subsequently reduced to 25 and 5 a.u., the measured 1/K decreased to 43.10 and 42.90, respectively. This confirmed that the LSCI-derived index followed both the increasing and decreasing phases of the imposed flow modulation. Linear regression (Fig. 5b) showed a strong positive relationship between pump speed and the LSCI-derived relative flow index, as: $\frac{1}{K} = 0.0396 \times \text{Speed} + 42.62$; and $R^2 = 0.9518$.

The high coefficient of determination indicated that the LSCI module was sensitive to controlled flow changes over the tested range. A small difference between the increasing and decreasing flow stages was observed, which may reflect residual flow, fluid inertia, or hysteresis within the tubing system. Overall, the phantom validation confirmed that the newly integrated LSCI subsystem could reliably track relative flow changes, supporting the use of rBFI as a perfusion-sensitive parameter in the subsequent rat spinal cord experiments.

*3.2 Exemplary multimodal maps from rat spinal cord imaging*

Exemplary multimodal maps were generated from rat spinal cord imaging to demonstrate the spatial outputs of the upgraded hNIR+ platform, as reported in Fig. 6 for one exemplary rat subject (others are reported in the Supplementary Materials). The 570 nm greyscale image was used as a high-contrast anatomical reference to visualise the exposed spinal cord surface and vascular structures, including the area of maximum confidence in dotted line (based on focusing and tissue curvature effect) for the choice of the ROIs (as explained in Sec 2.7). HSI-derived maps were generated for $\Delta HbO_2$, ΔHHb and ΔoxCCO, while LSCI provided rBFI maps.

The maps in Fig. 6 covered multiple physiological phases, including normoxia under 21% $O_2$, repeated hypoxia under 10% inspired $O_2$, and recovery under 21% $O_2$. During the 10% $O_2$ hypoxic phases, the HSI-derived haemoglobin maps showed generally the expected oxygenation response: $\Delta HbO_2$ decreased relative to normoxia, while ΔHHb increased. The partial reversal of these haemoglobin changes during recovery under 21% $O_2$ further confirmed that the measured signals were responsive to the respiratory gas challenge. The ΔoxCCO maps also changed during the hypoxic phases, showing a general decrease relative to the normoxic baseline.

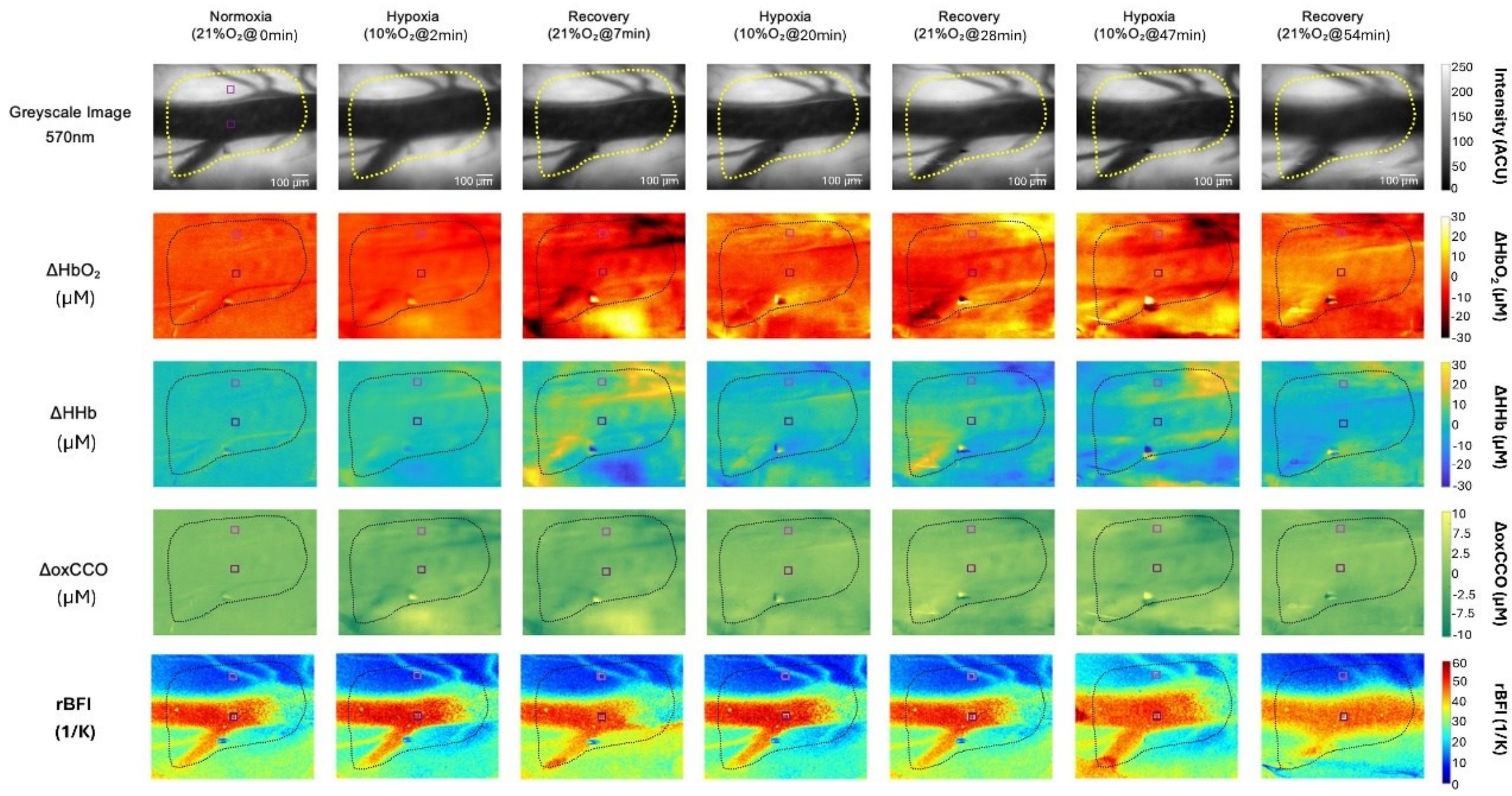

*Figure 6. Exemplary multimodal maps of the exposed rat spinal cord during normoxia, repeated 10% $O_2$ hypoxia and recovery under 21% $O_2$. The 570 nm greyscale image provides the anatomical reference, including the area of confidence (dotted line) for the maximum image quality based on focusing and tissue curvature effect. HSI-derived maps show $\Delta HbO_2$, ΔHHb and ΔoxCCO, while LSCI-derived maps show rBFI. Coloured square markers indicate the manually selected intravascular and extravascular ROIs, each with a size of 100 × 100 pixels, used for subsequent ROI-based time-course analysis (Fig. 7 and 8). The maps illustrate spatial changes in oxygenation, oxidative metabolism and perfusion responses across repeated respiratory gas challenges.*

Furthermore, across the repeated hypoxia and recovery phases, rBFI changes were also consistent with expected physiology, showing in most cases initial, compensatory increases during hypoxia followed by decrease, as the hypoxic phase extended, and then subsequent returns to quasi-baseline. These changes were most evident in vessel-containing regions and were spatially heterogeneous across the field of view, between extravascular and intravascular regions, with extravascular regions showing more limited changes and, overall, more immediate decreases compared to baseline.

*3.3 ROI-based time-course analysis across selected rat spinal cord experiments*

ROI-based time-course analysis was performed to evaluate the dynamic multimodal responses of the upgraded hNIR+ platform during respiratory gas challenges. Three independent rat spinal cord experiments were conducted in total, as reported in Fig. 7 (for the representative animal whose map were shown in Fig. 6) and 8 (for the other two animal subjects). The coloured square markers in Fig. 6 indicate an example of the manually selected, intravascular (specifically, the major dorsal vein) and extravascular ROIs used for subsequent quantitative analysis (the ROIs used for the other two rats are indicated in Fig. S1 and S2, respectively, in the Supplementary Materials). Each ROI was defined as a 100 × 100-pixel region, centred upon the desired tissue compartment (one extravascular, and one on the major dorsal vein).

In the first representative experiment (Fig. 7a), corresponding to the same animal dataset as the maps shown in Fig. 6, four repeated 10% $O_2$ hypoxic challenges produced $\Delta HbO_2$ decreases of approximately −6.17 to −13.04 μM and ΔHHb increases of approximately 6.96 to 13.67 μM in the main vasculature ROI. Only the first hypoxic phase produced an inverse response which may have been caused by the short duration of the oxygen challenge (less than 3 minutes). Extravascular responses were generally smaller in magnitude, by approximately 20 to 40%.

During the later severe 7.5% $O_2$ hypoxic phase, the response became substantially stronger, with $\Delta HbO_2$ decreasing to nearly −30.11 μM and ΔHHb increasing to nearly 30.92 μM in the vasculature ROI (about 75% less in the extravascular ROI). This terminal severe hypoxic phase did not represent a stable physiological challenge, but rather a transition towards circulatory failure and animal death. Therefore, unlike the repeated 10% $O_2$ hypoxic phases, the severe 7.5% $O_2$ phase showed a rapid deterioration rather than a new equilibrium response.

Moreover, during the majority of the hypoxic periods in the same animal, ΔoxCCO showed a decrease relative to baseline in the extravascular ROI, with typical changes ranging from -1.2 to -4.6 μM. During the terminal 7.5% $O_2$ phase in Fig. 7, the decrease in oxCCO became more pronounced (maximum of -6.7 μM), consistent with severe impairment of mitochondrial oxidative metabolism under critically reduced oxygen delivery.

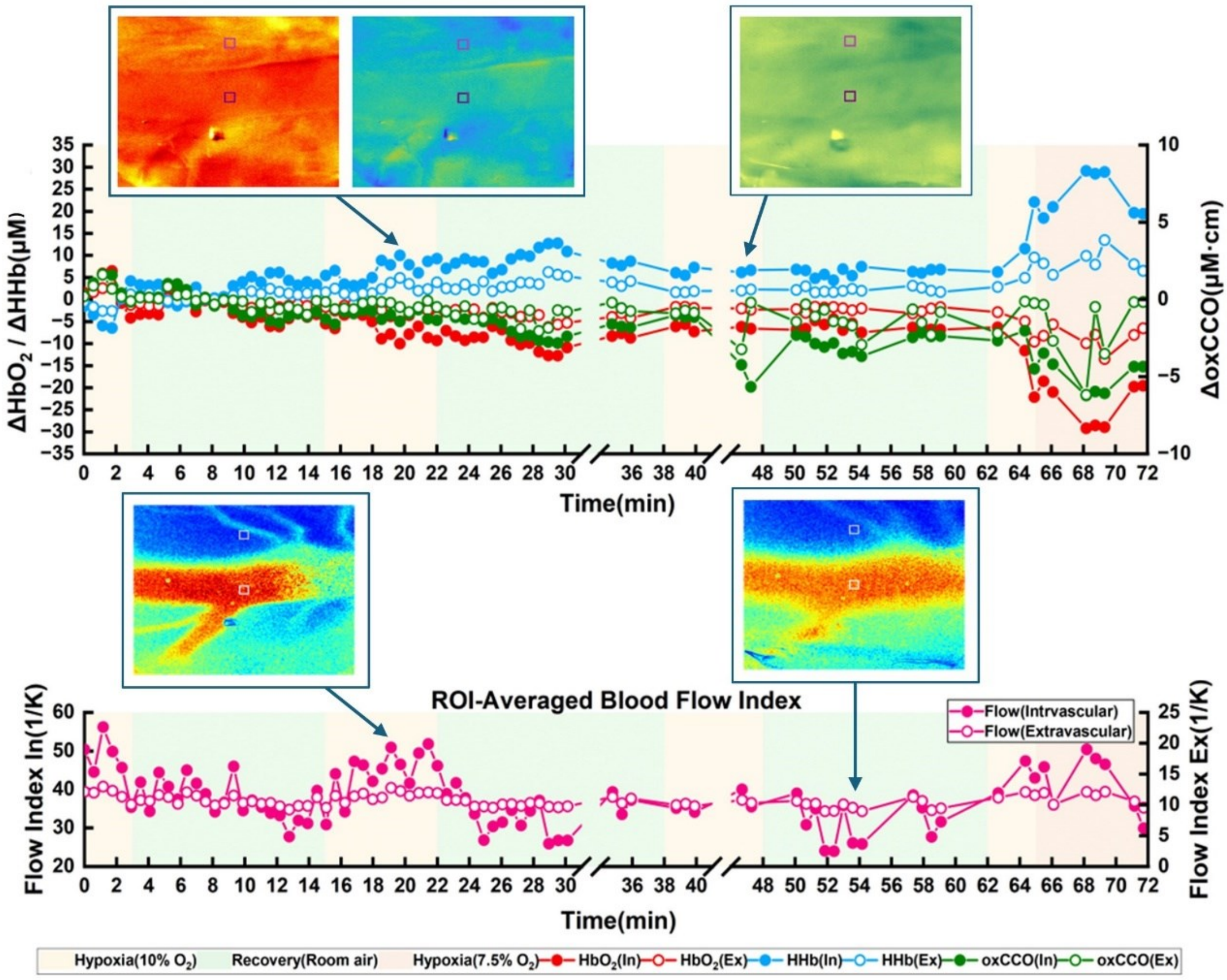

*Figure 7. ROI-based multimodal time-course analysis from the same maps on rat spinal cord for the subject in Fig. 6. Each dataset includes HSI-derived $HbO_2$, HHb and oxCCO signals, and LSCI-derived relative blood-flow index. Filled markers indicate intravascular ROIs and open markers indicate extravascular ROIs for all the traces. Shaded regions indicate respiratory gas challenge phases, including 10% $O_2$ hypoxia, room-air recovery, and severe hypoxia (7.5% $O_2$) with corresponding durations.*

The intravascular flow index typically ranged from approximately 35 to 58 1/K during the hypoxic phases, with the maximum reached during the latter severe hypoxia, whereas the extravascular flow index was smaller, typically ranging from approximately 5 to 20 1/K and showing much flatter responses over time. These differences support the separate analysis of vascular and extravascular regions, since averaging over the whole field of view could obscure compartment-specific flow dynamics. During the terminal severe hypoxic phase, the blood-flow response no longer reflected a compensatory or steady-state vascular regulation, but instead indicated failure of perfusion support.

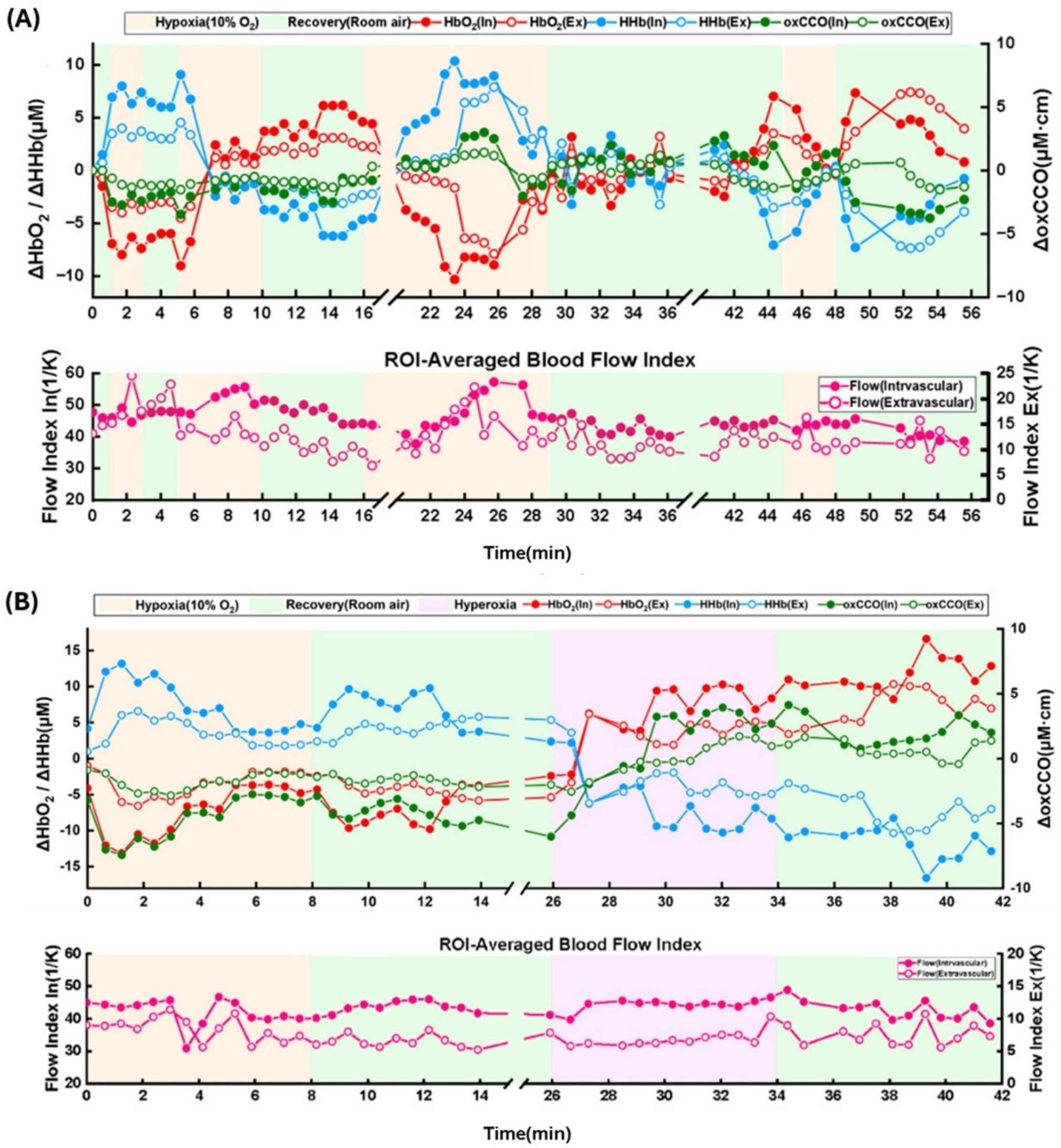

*Figure 8. ROI-based multimodal time-course analysis from the other two rat spinal cord experiments (A and B). Each dataset includes HSI-derived $HbO_2$, HHb and oxCCO signals, and LSCI-derived rBFI. Filled markers indicate intravascular ROIs and open markers indicate extravascular ROIs for all the traces. Shaded regions indicate respiratory gas challenge phases, including 10% $O_2$ hypoxia, room-air recovery, as well as hyperoxia for subject C. The results illustrate dynamic and region-dependent oxygenation, perfusion and metabolic responses across representative experiments.*

Similar responses are reported also for the other two animals (Fig. 8b and 8c), with ΔHHb generally increasing during the hypoxic phases 10% $O_2$ at up to a maximum of 10-15 μM and $\Delta HbO_2$ decreasing to a maximum of -10-12 μM. ΔoxCCo also reported decreases during the hypoxic phases in the majority of cases (up to a maximum of -5.12 μM), yet one animal also showed a partial increases during reduced oxygenation (maximum of 2.61 μM), highlighting potential heterogeneity in the metabolic shifts to the same type of oxygenation challenge. However, this remained an isolated case across all animals, for a total of eight hypoxic phases. rBFI trends were also consistent with the compensatory response seen in the first animal, yet with more substantial extravascular changes for the other two subjects. This may suggest differences in vascular compartmentalisation and overall peripheral perfusion.

Noteworthy, one dataset also included a hyperoxic phase at 100% $O_2$ (Fig 8b). During hyperoxia, the haemoglobin traces showed the expected response, with increased $\Delta HbO_2$ (maximum of 11.3 μM) and reduced ΔHHb (maximum of -10.15 μM). rBFI showed only moderate increase (7-8 1/K) while the oxCCO signal also increased during this phase (maximum of 5.22 μM).

## 4. Discussion

Overall, the representative maps preliminarily demonstrated that the upgraded hNIR+ platform can provide spatially resolved information on oxygenation, oxidative metabolism and relative perfusion from the same exposed rat spinal cord region. The parameter-specific responses observed during repeated 10% $O_2$ hypoxia and 21% $O_2$ recovery are generally consistent with the expected physiological effects of systemic hypoxia and support the proof-of-concept capability of hNIR+ to resolve coupled haemodynamic and metabolic changes *in vivo*.

Furthermore, the selected ROI-based time courses demonstrated that the upgraded hNIR+ system can track dynamic changes in haemoglobin oxygenation, CCO redox states, and relative blood flow during respiratory challenges. The combined analysis of $\Delta HbO_2$, ΔHHb, ΔoxCCO, and rBFI illustrated the added value of integrating LSCI-derived perfusion information with HSI-derived oxygenation and metabolic measurements in the same exposed spinal cord preparation. Taken together, the results suggest that hypoxia, hyperoxia and recovery cannot be fully described by a single parameter. $HbO_2$ and HHb indicate oxygenation changes, oxCCO provides metabolic redox information, and rBFI reflects relative perfusion. The differences among these signals demonstrate why multimodal imaging as provided by hNIR+, for the first time to this level of comprehensiveness and detail, is critically valuable for studying exposed spinal cord physiology, as well as similar tissue: it enables oxygen supply, oxygenation, perfusion, and metabolism-related information to be interpreted together rather than in isolation.

### *4.1 Added value of integrating LSCI into the hNIR+ platform*

The main aim of this study was to upgrade the existing hNIR platform to the new hNIR+ by integrating laser speckle contrast imaging, thereby adding perfusion-sensitive information to the original hyperspectral measurements of haemoglobin oxygenation and CCO redox state. The original hNIR system was able to provide spatial maps of $HbO_2$, HHb, and oxCCO [7], which are valuable for assessing oxygenation and oxidative metabolism in exposed neural tissue. However, these measurements alone cannot directly determine whether the observed oxygenation or metabolic changes are associated with altered blood flow. The addition of LSCI addressed this limitation by providing a relative blood-flow index from the same exposed spinal cord region. Similar multimodal optical systems have shown that combining LSCI with hyperspectral or multispectral imaging can provide complementary information on perfusion and oxygenation, improving tissue assessment compared with single-modality imaging [20, 21]. However, hNIR+ is the first (to our knowledge) to provide simultaneous information on haemoglobin, blood flow, as well as mitochondrial metabolism.

The blood-flow phantom experiment demonstrated that the newly integrated LSCI module was sensitive to controlled flow changes. The LSCI-derived inverse speckle contrast index increased with pump-controlled flow and showed a strong positive linear relationship with the imposed flow setting. This result confirmed that the LSCI subsystem could provide a reliable relative flow-sensitive measurement under controlled conditions before being applied to *in vivo* rat spinal cord imaging. This validation step was important because LSCI measurements are influenced by exposure time, scattering properties, optical geometry, and model assumptions, and are therefore more appropriately interpreted as relative perfusion indices rather than absolute blood-flow velocities.

### *4.2 Physiological interpretation of oxygenation, perfusion and metabolic responses*

The multimodal responses observed during the rat spinal cord experiments reflected different but complementary aspects of tissue physiology. The HSI-derived haemoglobin signals primarily described changes in tissue oxygenation and deoxygenation. During hypoxic challenges, the reduction in inspired oxygen concentration would be expected to decrease oxygen availability, leading to a decrease in $HbO_2$ and an increase in HHb, as correctly estimated with hNIR+. Similarly, during hyperoxia, $HbO_2$ and HHb had instead opposite responses to the increase in oxygen supply. The partial recovery of all these signals after the return to normoxic conditions suggested that the haemoglobin changes were responsive to the respiratory gas manipulation. However, the magnitude and temporal pattern of recovery differed between experiments, indicating that the spinal cord response may depend on local vascular condition, tissue state, and experimental variability.

The oxCCO signal provided additional information related to mitochondrial redox state and oxidative metabolism. During hypoxia, oxCCO was reported to decreases with magnitude according to severity of the oxygen flow reduction, while during hyperoxia it showed a general increase. Both these responses were physiologically expected because reduced oxygen delivery can limit mitochondrial oxidative metabolism and shift CCO towards a more reduced state (redCCO). Importantly, the oxCCO response was not spatially identical to the haemoglobin response. This difference supports the value of measuring mitochondrial redox state in addition to haemoglobin oxygenation, because haemoglobin signals primarily reflect vascular oxygenation, whereas oxCCO provides information related to extravascular cellular oxidative metabolism [4, 8].

Furthermore, the temporal behaviour of oxCCO did not always simply mirror the haemoglobin signals, which is physiologically important. Haemoglobin signals mainly reflect oxygen transport and vascular oxygenation, whereas oxCCO is more closely associated with cellular oxygen utilization and mitochondrial oxidative state. Therefore, a mismatch between haemoglobin recovery and oxCCO recovery may suggest that restoration of oxygen delivery does not necessarily imply immediate restoration of mitochondrial metabolic status [8]. This supports the value of including oxCCO in addition to $HbO_2$ and HHb when assessing spinal cord responses to hypoxia and recovery.

The LSCI-derived rBFI added a further dimension by showing relative perfusion changes. The vascular ROIs generally showed larger blood-flow-related variations than the extravascular ROIs, which is consistent with the higher flow contribution from visible vessels. The extravascular signal was smaller in amplitude but still useful, as it may better reflect tissue-level perfusion changes in regions with reduced large-vessel contribution. This difference between vascular and extravascular responses highlighted the need to analyse these compartments separately, rather than relying on a single averaged blood-flow trace over the entire field of view. This behaviour was expected because local vascular responses (for example to hypoxia) may vary between large vessels (primarily veins for the study) and surrounding extravascular tissue regions (microvasculature and capillaries).

Finally, it is also noteworthy to mention how the reconstructed maps generally express also some localised heterogeneity in the responses, which may be highlighted further with a different choice of the ROI used for the analysis. We have primarily looked at larger, systemic regions where physiological responses reflected theoretical changes based on literature, as a mean to provide preliminary proof-of-concept of the hNIR+ system. Further studies could investigate in depth this phenomenon. Nonetheless, the finding indeed confirms how the spatial mapping information does provide an additional value to the retrieved information that a point-based, only temporal analysis could not.

*4.3 Limitations and future improvements*

Although the upgraded hNIR+ platform demonstrated the feasibility of multimodal spinal cord monitoring, several limitations should be considered. First, the LSCI-derived rBFI was interpreted as a relative perfusion-sensitive index rather than an absolute blood-flow velocity. Speckle contrast is influenced by exposure time, static scattering, optical geometry, tissue optical properties, and motion artefacts. Therefore, the rBFI values in this study were mainly used to compare temporal changes within the same experiment, rather than to quantify absolute perfusion.

Second, the current iteration of the system used sequential rather than fully simultaneous acquisition. This approach reduced optical crosstalk between hyperspectral and speckle illumination, but it also means that very rapid physiological changes may not be captured at exactly the same time by the two modalities. Future improvements could include faster wavelength tuning, synchronized acquisition control, improved illumination switching, and more robust image registration to increase temporal accuracy [32].

A third limitation concerns image quality, primarily related to focusing. In the experiments, we deliberately indicated areas of confidence where the image focus was stable enough to provide satisfactory data quality. Outside this area, the image suffered from defocusing and blurring caused by curvature and geometry of the target tissue. Such issue is related to the limited depth of field of the reflective objective of the hNIR+ system, which was selected to provide the desired mesoscopic field-of-view but also guarantee achromatic aberration correction across the whole visible and NIR range. The limited depth of field for this application is currently a technological bottleneck and a trade-off with spatial resolution and spectral stability, that may be improved in the future with better and more suited optics.

Finally, the present study was designed as a characterization, validation, and proof-of-concept demonstration. The number of *in vivo* experiments was limited for the specific scope, and the physiological responses varied between independent datasets. Future work should include a larger animal cohort, improved control of physiological variables, absolute or semi-quantitative flow calibration, and more advanced models for estimating oxygen extraction and oxygen metabolism. Further validation in spinal cord injury, ischemia, or surgical monitoring models would also help evaluate the translational potential of the upgraded hNIR+ platform.

## 5. Conclusion

The study presented here described the upgrade of the hNIR imaging platform, a HSI-only instrument, through the integration of LSCI, thus leading to the creation of the new, multimodal, hNIR+ system. The upgraded platform enables simultaneous assessment of haemoglobin oxygenation, CCO redox state, and relative blood flow, in the exposed rat spinal cord. Phantom and *in vivo* experiments demonstrated sensitivity to flow changes and the ability to map and monitor physiological responses during various phases of normoxia, hypoxia, hyperoxia and subsequent recoveries. By combining oxygenation, metabolic, and perfusion measurements, hNIR+ provides a more comprehensive assessment of haemodynamic-metabolic coupling, helping distinguish changes in oxygen delivery, extraction, and utilisation.

Overall, the integration of LSCI extends hNIR from a system that primarily maps oxygenation and oxidative metabolism into a multimodal platform capable of assessing oxygenation, mitochondrial redox changes, and relative perfusion simultaneously and within the same experimental framework. This is, to our knowledge, the first time such a comprehensive optical instrument and capacity have ever been presented and validated *in vivo*.

The hNIR+ system represents a powerful tool for preclinical monitoring of spinal cord function and could be extended in the future to applications in the brain, retina, heart, and other tissues, enabling quantitative investigation of oxygenation, perfusion, and metabolism in both healthy and pathological conditions.

**Acknowledgements**

LG, JW, FL, and IL are supported by: (1) the HyperProbe project, as part of the European Union’s Horizon Europe research and innovation programme under grant agreement No 101071040 and of the UKRI grant number 10048387; (2) the fastMOT project, as part of the EU’s HORIZON EUROPE programme under grant agreement number 101099291 and of the UK Research and Innovation (UKRI) under the UK government’s Horizon Europe funding guarantee (grant number 10063660): and (3) the Medical Research Council (MRC) grant RE-SPECT4NEURODEVELOPMENT under grant agreement number EP/W035154/1. AGY and EG are supported by the International Progressive Multiple Sclerosis Alliance.